# Short cycle pulse sequence for dynamical decoupling of local fields and dipole-dipole interactions

S.A.Moiseev[1,2][*], and V.A.Skrebnev[2][**]

[1] Quantum Center, Kazan National Research Technical University,
10 K. Marx, Kazan, 420111, Russia

[2] Zavoisky Physical-Technical Institute of the Russian Academy of Sciences,
10/7 Sibirsky Trakt, Kazan, 420029, Russia

E-mails: * samoi@yandex.ru, ** vskrebnev@mail.ru

We propose a new pulse sequence for dynamical averaging of the dipole-dipole interactions and inhomogeneities of the magnetic fields in the nuclear spin system. The sequence contains a short cycle of the periodic resonant pulse excitation that offers new possibilities for implementing the long-lived multi-qubit quantum memory on the condensed spin ensembles that are so important for the construction of a universal quantum computer and long-distance quantum communications.

*Introduction.* The development of a quantum computer and long-distance optical quantum communications stimulates the creation of the microwave and optical quantum memories (QM) [1-3]. Herein, the elaboration of the universal QM capable of operating with one- and multi-qubit states of light is of great interest. Recently a significant progress has been achieved when using macroscopic coherent atomic ensembles [4,5,6,7,8]. A series of impressive experimental results on the basis of the photon echo technique [7,9] was obtained that made it possible to achieve the record quantum efficiency of 67% for the quantum storage in the solid state [10] and 87% in the atomic gas [11]. Using single QM cell for the storage of single photon fields [12,13], several tens of photons [14], and more than thousand of light pulses [15] that is necessary for the implementation of a quantum repeater and a universal quantum computer has been also demonstrated on the basis of this technique.

Solid-state media, amongst which promising results on long-lived storage were achieved on ensembles of rare-earth ions in organic crystals [7,10,14,15], NV centers in diamond [16,17,18,19,20], phosphorus ions in silicon [21,22], are of great interest for applications. At the same time the higher quantum efficiency and longer lifetime of the QMs are necessary for the implementation in practice that push the solution of a crucial series of new problems in solid state physics, spectroscopy and controlled quantum dynamics of multi-particle systems. Here, great expectations on the increase in the lifetime of the quantum memory in such media are associated with transfer of the mapped flying photon qubits to the long-lived electron and nuclei spin states [23,24] which is accompanied by the development of the experimental methods providing active dynamic suppression of the decoherence (the so-called dynamical decoupling (DD)) in the interacting nuclei and electron spins [25,26,27].

Originally, the DD methods were proposed during the development of the high-resolution multi-pulse nuclear magnetic resonance (NMR) methods [28,29]. The creation of QM and quantum computer requires searching for new efficient DD methods of the decoherence suppression which could work successfully with the quantum systems used for quantum processing. Recently new DD pulse sequences were proposed for the suppression of the decoherence effect in the qubit evolution caused by the fast fluctuating spin-bath and pulse imperfections [30]. However, the demonstrated efficient multi-qubit QM assumes the presence of the inhomogeneous broadening of the resonance line of atoms [9,15,27,31]. Such condensed multi-atomic systems suffer from the long-distance dipole-dipole interaction in the condensed media that strongly affects the atomic coherence. The inhomogeneous broadening and dipole-dipole interactions are the two main sources of the electron and nuclei decoherence even at lowest temperature and should be efficiently suppressed in any multi-mode QM.

Different sequences of the resonance radio-frequency pulses are used now in order to increase the coherence time of the nuclear spin systems. The Carr-Purcell pulse sequence and its modification, the Carr-Purcell-Meiboom-Gill (CPMG) sequence, are used for suppressing the effect of the inhomogeneities of the magnetic field [32,33]. In turn, the WAHUHA [34] and MREV8 [35] (see also [36]) sequences are used for averaging the dipole-dipole interactions. One has to apply the composite variants of the pulse sequences mentioned above in order to more efficiently weaken the decoherence in the nuclear spin qubits due to the simultaneous presence of the dipole-dipole interactions, as well as the local inhomogeneities and the magnetic field drift [19,36]. Unfortunately, such composite sequences contain a large number of pulses that increases the time of the cycle used for averaging the undesirable interactions. The large number of pulses in the cycle limits the applicability of the sequence imposing restrictions on the minimum initial decoherence time of the considered systems, as well as inevitably leads to the increase in the errors caused by the imperfection of the used pulses [28].

In this work we propose a new pulse sequence containing a much less number of pulses in the cycle than that in the composite sequences, moreover the proposed sequence takes into account in first principles the microscopic Hamiltonian of the analyzed multi-atomic system. The elaborated pulse-sequence makes it possible to simultaneously suppress the decoherence in the system of qubits, which is due to the effects of the inhomogeneities of the magnetic field and spin-spin interaction. The proposed technique demonstrates a new possibility for the efficient increase in the lifetime for multi-qubit QMs in the concentrated multi-atomic ensembles. Finally, we discuss the advantages of using this sequence for the implementation of the long-lived QMs and potential applicability of this technique in practice.

*Pulse Sequence.* The use of nuclear spins, on which the promising results on the storage time of quantum information at helium [21,27] and room [19,22] temperatures were obtained to date, is particularly perspective for achieving the longest lifetime. However, it is necessary to use a large number of nuclear spins in order to write the quantum state of many qubits in a single quantum memory cell. Let a quantum state given by the density matrix $\rho(0)$ be excited at the initial instant of time in a system of nuclear spins. In particular, this quantum state can be created with the use of one of the protocols of the photon (spin) echo [7,9,10,11,31]. To suppress the decoherence effects in the considered system of spins caused by the presence of the magnetic field inhomogeneities and spin-spin interactions and, accordingly, to increase the lifetime of the quantum memory, we propose to use the sequence of the radio-frequency pulses shown in Fig. 1.

We consider the pulses of the radio-frequency field as short $\delta-$ pulses. Let us present the main formulas for the description of the spin evolution under the action of the proposed pulse sequence [28]. The total Hamiltonian of this system has the form:

$$H = H_1(t) + H_{in}, \qquad (1)$$

here $H_{in}$ and $H_1(t)$ are Hamiltonians of the internal interactions and the interaction with external radio-frequency field of the considered multi-spin system.

The density matrix of the system at the arbitrary instant of time t is written as follows:

$$\rho(t) = U(t)\rho(0)U^{-1}(t), \qquad (2)$$

where the evolution operator $U(t) = U_1(t)U_{in}(t)$,

$$U_1(t) = Texp\left[-i\int_0^t H_1(t')dt'\right], \qquad (3)$$

$$U_{in}(t) = Texp[-i\int_0^t \widetilde{H}_{in}(t')dt'], \qquad (4)$$

here T is the Dyson time-ordering operator,

$$\widetilde{H}_{in}(t) = U_1^{-1}(t)H_{in}U_1(t). \qquad (5)$$

If $H_1(t + t_c) = H_1(t)$ and $U_1(t + t_c) = U_1(t)$, the time $t_c$ is called the cycle time. In this case with allowance for $U_1(0) = 1$, we obtain $U_1(t_c) = 1$, and also $\widetilde{H}_{in}(t) = \widetilde{H}_{in}(t + t_c)$ and $U_{in}(Nt_c) = [U_{in}(t_c)]^N$.

Let us use the Magnus expansion [28]:

$$U_{in}(t_c) = Texp[-i\int_0^{t_c} \widetilde{H}_{in}(t')dt'] = exp[-i(\bar{H} + \bar{H}^{(1)} + \bar{H}^{(2)} + \cdots)t_c], \qquad (6)$$

where

$$\bar{H} = \frac{1}{t_c}\int_0^{t_c} \widetilde{H}_{in}(t)dt, \qquad (7)$$

$$\bar{H}^{(1)} = \frac{-i}{2t_c}\int_0^{t_c} dt_2 \int_0^{t_2} dt_1 [\widetilde{H}_{in}(t_2), \widetilde{H}_{in}(t_1)], \qquad (8)$$

etc.

In the coordinate system rotating with the resonance frequency the Hamiltonian of the system of nuclear spins in the inhomogeneous magnetic field has the form [37]:

$$H_{in} = \sum_i \Delta_i I_z^i + \sum_{i>j} a_{i,j}(3 I_z^i I_z^j - I^i I^j), \qquad (9)$$

where $\hat{I}_{x,y,z}^i$ are spin operators, $\Delta_i$ is the frequency detuning of the i-th nuclear spin, $a_{i,j} = \frac{1}{4}\gamma^2 \hbar^2 \frac{(1-3\cos\theta_{i,j})}{r_{i,j}^3}$ is the constant of the spin-spin interaction, $\gamma$ is the nuclear gyromagnetic ratio, $\hbar$ is the Planck constant, $\vec{r}_{i,j}$ is the distance between i-th and j-th spins, $\theta_{i,j}$ is the angle between the external magnetic field $\vec{H}_z$ and $\vec{r}_{i,j}$.

To describe the action of the pulse sequence on the spin system, it is convenient to select a pair of interacting spins with the internal Hamiltonian:

$$H_{12} = \Delta_1 I_z^1 + \Delta_2 I_z^2 + a(3 I_z^1 I_z^2 - I^1 I^2). \qquad (10)$$

Ignoring the correction terms $\bar{H}^{(k)}$ in the exponent (6) [28], we limit ourselves to finding the first term $\bar{H}$, which is called the average Hamiltonian.

Under the action of the pulse sequence Fig. 1 according to formulas (3) and (5) we have the following value for the operators $U_1(t)$ and $\widetilde{H}_{12}(t) = U_1^{-1}(t) H_{12} U_1(t)$ in the following six time intervals between $t = 0$ and $t = t_c = 6\tau$:

1) $U_1(0 - \tau) = 1$,
$\widetilde{H}_{12}(0 - \tau) = \Delta_1 I_z^1 + \Delta_2 I_z^2 + a(3 I_z^1 I_z^2 - I^1 I^2)$.

2) $U_1(\tau - 2\tau) = \exp(-i\frac{\pi}{2} I_x)$,
$\widetilde{H}_{12}(t - 2\tau) = \Delta_1 I_y^1 + \Delta_2 I_y^2 + a(3 I_y^1 I_y^2 - I^1 I^2)$.

3) $U_1(2\tau - 3\tau) = \exp(i\frac{\pi}{2} I_y)\exp(-i\frac{\pi}{2} I_x)$,
$\widetilde{H}_{12}(2t - 3\tau) = \Delta_1 I_x^1 + \Delta_2 I_x^2 + a(3 I_x^1 I_x^2 - I^1 I^2)$.

4) $U_1(3\tau - 4\tau) = \exp(-i\frac{\pi}{2} I_y)\exp(-i\frac{\pi}{2} I_x)$,
$\widetilde{H}_{12}(3t - 4\tau) = -\Delta_1 I_x^1 - \Delta_2 I_x^2 + a(3 I_x^1 I_x^2 - I^1 I^2)$.

5) $U_1(4\tau - 5\tau) = \exp(i\frac{\pi}{2} I_x)$,
$\widetilde{H}_{12}(4t - 5\tau) = -\Delta_1 I_y^1 - \Delta_2 I_y^2 + a(3 I_y^1 I_y^2 - I^1 I^2)$.

6) $U_1(5\tau - 6\tau) = \exp(i\pi I_x)$,
$\widetilde{H}_{12}(5t - 6\tau) = -\Delta_1 I_z^1 - \Delta_2 I_z^2 + a(3 I_z^1 I_z^2 - I^1 I^2)$.

For the fulfillment of the cyclicity condition it is necessary to apply the pulse $\pi_{-x}$ at the moment of time $t = 6\tau$. However, it is possible to ignore the contribution to the average Hamiltonian $\bar{H}_{12}$ from the impact of the $\pi_{-x}$ − pulse.

Finding the average Hamiltonian $\bar{H}_{12}$ for the total pulse sequence (see Fig.1) during the time of the cycle we obtain

$$\bar{H}_{12} = \int_0^{t_c} \widetilde{H_{12}}(t) dt = 0. \tag{11}$$

It is obvious that the average Hamiltonian of the whole system of spins is also zero. We note that the pulses 4 and 5 should be located as close as possible. Namely, the time delay between these two pulses should be much less than the free induction decay time that can be easily implemented experimentally. Otherwise the proposed DD sequence will not work efficiently. Thus, the action of the pulse sequence leads to the simultaneous suppression of the effect of the inhomogeneities of the magnetic field and the dipole-dipole interactions on the quantum state of the nuclear spin system that provides the considerable increase in the decoherence time of the nuclear spins. In principle, such pulse sequence can be also implemented for the electron spin ensembles. The creation of a shorter DD pulse sequence providing the effective suppression of the decoherence sources is an open question. It is worth noting that the approximation of the average Hamiltonian is correct for the sufficiently short time of the sequence cycle $t_c \ll t_d$, where $t_d$ is the initial decoherence time of the considered system of nuclear spins. The number of radio-frequency pulses in the cycle in the composite sequence used recently in [19] is 34. There are seven radio-frequency pulses in the cycle of the sequence we propose, therefore it is much easier to provide the fulfillment of the condition $t_c \ll t_d$ for the further increase in the QM lifetime. This pulse sequence will also effectively average out the drift of the spin frequency in the considered system of nuclei, if the drift rate does not exceed $t_c^{-1}$. In addition, the negative effect of the pulse imperfection on the implementation of the long-lived quantum memory in the system of nuclear spins is suppressed with the decrease in the number of pulses that promises additional advantages in experimental applications.

In conclusion, we note that the new pulse sequence has been elaborated for the multi-atomic system characterized by inhomogeneously broadened resonant line and microscopic Hamiltonian of dipole-dipole interactions that is applicable for arbitrary electron and nuclei multi-spin systems. To the best of our knowledge, there are no longer (multi-pulse) DD sequences which could handle the studied problem more efficiently than the sequence proposed in this work. Since the analyzed approach is based on the first principles consideration of the multi-particle system, we believe that it could be successfully applied for efficient dynamical decoupling of various resonant systems characterized by similar microscopic Hamiltonians. By taking into account highly developed multi-pulse NMR experimental technique, we note that the proposed pulse sequence will be a useful instrument in experiments to achieve the longer lifetime of the multi-qubit QM on the atomic ensembles in condensed media that are vital for the construction of the universal quantum computer and long-distance quantum communications. The proposed approach could be useful for the elaboration of new effective DD sequences.

This work has been financially supported by the Russian Scientic Fund through the grant no. 14-12-01333.

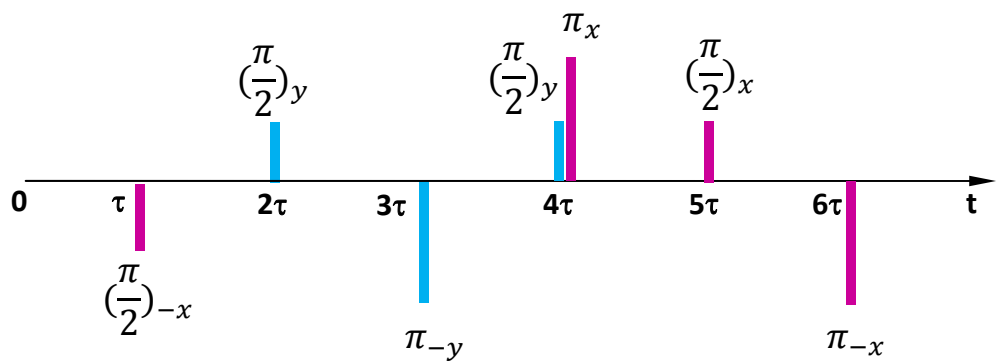

Fig.1.

**Caption for the Fig.1** (color on line): Sequence of 6 rf- delta pulses separated by equal time delays τ; the magnetic fields of these pulses are oriented along x(-x) axis - brawn up(dawn) color lines or along y(-y) axis – blue up(down) lines; pulses 4 and 5 are closely spaced to each other.